\documentclass[twocolumn,aps,prb,citeautoscript,superscriptaddress]{revtex4} 
\usepackage{graphicx}
\usepackage{amsmath}
\begin{document}

\title{The electrostatic potential profile along a biased molecular wire: A model quantum mechanical calculation}

\author{St\'ephane Pleutin}
\author{Hermann Grabert}
\affiliation{Physikalisches Institut, Albert-Ludwigs-Universit\"at, 
Hermann-Herder-Stra{\ss}e 3, D-79104 Freiburg, Germany}
\author{Gert-Ludwig Ingold}
\affiliation{Institut f\"ur Physik, Universit\"at Augsburg, 
Universit\"atsstra{\ss}e 1, D-86135 Augsburg, Germany}
\author{Abraham Nitzan}
\affiliation{School of Chemistry, The Sackler Faculty of Science, Tel Aviv
University, 69978 Tel Aviv, Israel}

\date{\today}
\begin{abstract}
We study the electrostatic potential of a molecular wire bridging two metallic 
electrodes in the limit of weak contacts. With the use of a tight-binding model 
including a fully three-dimensional treatment of the electrostatics of the 
molecular junction, the potential is shown to be poorly screened, dropping 
mostly along the entire molecule. In addition, we observe pronounced Friedel 
oscillations that can be related to the breaking of electron-hole symmetry. Our 
results are in semi-quantitative agreement with recent state-of-the-art 
\textit{ab initio} calculations and point to the need of a three-dimensional
treatment to properly capture the behavior of the electrostatic potential. 
Based on these results, current-voltage curves are calculated within the 
Landauer formalism. It is shown that Coulomb interaction partially compensates 
the localization of the charges induced by the electric field and consequently 
tends to suppress zones of negative differential resistance. 
\end{abstract}
\maketitle

\section{Introduction}
Due to important technical progress, the field of molecular electronics, born 
in the mid 70s with the proposal of Aviram and Ratner to use single organic 
molecules as rectifiers \cite{aviram}, receives rapidly growing interest 
\cite{joachim,nitzan,chemphys}. Indeed, new fabrication methods and probes now 
enable individual molecules or small numbers of them to be connected to macroscopic 
electrodes \cite{joachim,reed,bourgoin,cui,karlsruhe}. Among these methods, one 
may cite, for instance, the break-junction technique 
\cite{bourgoin,reed,karlsruhe} and the use of a conducting
Atomic-Force-Microscope (AFM) to contact molecules absorbed on a metallic 
surface \cite{cui}.

On the theoretical side, the problem posed by these experimental works is 
highly challenging. We are facing a non-equilibrium many-body problem where, 
moreover, the coupling to a phonon bath may also be of importance. Up to now, 
most of the studies have focussed on the coherent regime and the Landauer 
approach has been employed to obtain the conductance from \textit{ab initio} or 
semi-empirical models \cite{nitzan}. Important inelastic processes were 
included only within simple models \cite{petrov,lehmann,segal,pastawski,ness} and 
much further progress is needed before one may hope to reach a satisfactory 
understanding of the problem.

The exact number of molecules contacted by the leads remains for a 
large part uncontrolled in the experimental set-ups cited above 
\cite{joachim,karlsruhe}. In theoretical modelling it is convenient to
assume that a single organic molecule bridges two semi-infinite metallic 
electrodes (cf. Fig.~\ref{fig1}). Another important experimental aspect is
the fact that current-voltage characteristics are measured with applied 
voltages up to a few volts, values 
which bring us well away from the linear regime.

In this context, a central question concerns the electrostatic potential 
profile of a biased molecular wire. The importance of this issue was first 
demonstrated by Datta and coworkers \cite{datta,tian}. Using semi-empirical 
models, they have shown that different choices of the electrostatic profile 
have a profound effect on the current-voltage characteristics of a molecular 
junction. For instance, the transport properties are strongly modified 
depending on whether the potential drop occurs at the interface between 
molecule and electrode or along the molecular wire. In fact, it is natural 
to assume that even the details of the potential shape have a considerable 
effect on molecular conductance.

Recently, a few works along that line have been reported 
\cite{mujica1,lang,damle}. They give us a rather ambiguous view of this 
fundamental problem. Model calculations involving self-consistent solutions of 
the coupled Poisson and Schr\"odinger equations suggest that the potential drop 
occurs mainly at the interface between the molecule and the electrodes 
\cite{mujica1}. Within the molecule, the electrostatic potential is then found 
to be essentially flat. Screening appears to be very efficient within this 
approach and the final conclusions are in agreement with some previous 
investigations \cite{datta,tian}. However, these model calculations involve a 
drastic approximation: instead of solving the full Poisson equation, the 
authors of Ref.~~\onlinecite{mujica1} have used a one-dimensional version of 
it. Implicitly, it is then assumed that the lateral dimensions of the molecule 
are much larger than the screening length. For quasi-one dimensional systems 
with lateral dimensions of the order of a few angstr{\"o}ms, such as the 
organic molecules used in recent experimental work, this approximation is 
clearly questionable \cite{nitzan2002}. Indeed, recent state-of-the-art \textit{ab initio} calculations on 
carbon and gold chains show a quite different picture \cite{lang,damle}. In 
these works, the potential drop occurs not only at the interface but rather 
along the entire molecule. Moreover, the local potential is found to display 
pronounced Friedel oscillations. Contrary to previous results, screening 
appears to be rather inefficient, even for metallic wires.

We are then left with two different pictures and it becomes clear that a full 
understanding of the electrostatic potential profile in biased molecular wires 
or metallic constrictions is still lacking. In this work, we re-address the 
problem following the approach by Mujica \textit{et al.}\ \cite{mujica1} We 
perform model calculations and solve the 
coupled set of two equations: the Poisson equation for the 
electrostatics and the Schr\"odinger equation for the electronic structure. 
In this respect, our work is similar to that of Mujica \textit{et al.}\ 
\cite{mujica1} However, the calculation is modified in ways that we believe to 
be essential. In particular, we treat the real three-dimensional Poisson 
equation. As recently discussed by us \cite{nitzan2002}, we expect that this 
proper handling of the electrostatic problem changes the qualitative behavior: 
The potential is poorly screened and falls off substantially along the
molecule. Indeed, the electrostatic potential profile of the model calculations 
presented here is in semi-quantitative agreement with the \textit{ab initio} 
results reported in Refs.~~\onlinecite{lang} and~~\onlinecite{damle}. However, while 
\textit{ab initio} calculations are involved and intricate enough to 
leave the underlying physics essentially obscure, our model includes only the 
ingredients necessary to capture the correct screening effects, and consequently our calculations are rather economic in time. We thus believe 
that our approach may help to gain further insight into the difficult problem 
of understanding transport through molecules, i.e.\ at a scale where quantum 
effects are prominent. A forthcoming work by Ghosh \textit{et al.}\ \cite{ghosh} reaches conclusions similar to those presented here.

In section~\ref{sec:model}, our model Hamiltonian is introduced. The electronic 
density in the absence of a bias potential is then studied in 
section~\ref{sec:nobias}, using exact diagonalization and Hartree-Fock 
calculations that are shown to agree reasonably well which each other. The 
electrostatic profile is then calculated in section~\ref{sec:bias} at the 
Hartree-Fock level.  Finally, the current-voltage characteristics are discussed 
in section~\ref{sec:iv}.

\section{Model Hamiltonian}
\label{sec:model}

\subsection{Coulomb interaction including image charges}

\begin{figure}
\includegraphics[width=\linewidth]{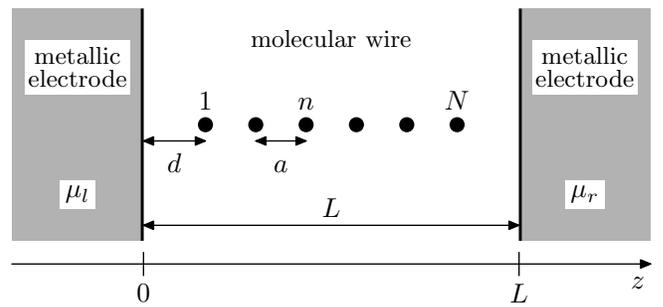}
\caption{Idealized molecular junction. A molecular wire, modelled by a finite 
one-dimensional lattice, bridges two metallic electrodes with surfaces assumed 
to be infinite planes. The tunnelling contacts, effective only at the two 
molecular end sites indicated by $1$ and $N$, are assumed to be weak. The 
chemical potentials in the left and right electrode are denoted by
$\mu_l$ and $\mu_r$, respectively.}
\label{fig1}
\end{figure}

The physical problem posed by a molecular wire between two infinite metallic 
reservoirs is far too complicated to be solved exactly and, to proceed, several 
approximations are necessary. First, we assume that the surfaces of the two 
electrodes are infinite planes (cf.\ Fig.~\ref{fig1}). Second, the molecule is 
assumed to be weakly connected to the metallic electrodes so that their 
chemical constitution is unimportant. This is certainly not fulfilled for some 
of the wires examined experimentally, with covalent bonds between molecule and 
electrode. Finally, we assume the characteristic time scale for electronic 
processes in the electrodes to be much shorter than the transit time of 
electrons in the wire. The electrodes can then be treated as equipotential 
surfaces, and the Schr\"odinger equation is solved under these potential boundary conditions. We are then within the same framework used in 
Ref.~~\onlinecite{mujica1}, but proceed differently. We first determine the 
Coulomb interaction potential which includes the image charges due to the 
metallic leads, keeping its three-dimensional character. Then, we solve the 
electronic wire problem.

The Coulomb interaction energy reads
\begin{equation}
W=\frac{1}{2}\int d^3\mathbf{r}\int d^3\mathbf{r'}\rho(\mathbf{r})\rho(\mathbf{r'})
\varphi(\mathbf{r},\mathbf{r'})
\end{equation}
where $\rho(\mathbf{r})$ is the charge density in the wire and
$\varphi(\mathbf{r}, \mathbf{r'})$ is the potential at point 
$\mathbf{r}=(x,y,z)$ produced by the charge located at point 
$\mathbf{r'}=(x',y',z')$. The Coulomb potential is the solution 
of the Poisson equation
\begin{equation}
\Delta_{\mathbf{r}}\varphi(\mathbf{r},\mathbf{r'})=
-4\pi U\delta(\mathbf{r}-\mathbf{r'})
\label{eq:poisson}
\end{equation}
where for convenience we have introduced a factor $U$ measuring the strength of 
the electron-electron interaction. In the absence of the metallic
electrodes the solution of (\ref{eq:poisson}) is given by the standard
Coulomb potential
\begin{equation}
\varphi_0(\mathbf{r},\mathbf{r'}) = 
\frac{U}{\vert\mathbf{r}-\mathbf{r'}\vert}\;.
\end{equation}

In the setup depicted in Fig.~\ref{fig1}, we require the potential in the 
absence of an external bias to vanish on the surfaces of the metallic 
electrodes, i.e.\ at $z=0$ and $z=L$. A solution of the Poisson equation with 
these particular boundary conditions is found using the standard method of 
image charges
\begin{equation}
\varphi(\mathbf{r},\mathbf{r'}) = \sum_{p=-\infty}^{+\infty}\left[
\varphi_0(\mathbf{r}+2pL \mathbf{\hat{z}},\mathbf{r'})
-\varphi_0(2pL \mathbf{\hat{z}}-\mathbf{r},\mathbf{r'})\right]
\label{coulomb}
\end{equation}
where $L$ is the distance between the two electrodes (cf.\ Fig.~\ref{fig1})
and $\mathbf{\hat{z}}$ the unit vector along the molecular axis.

$\varphi(\mathbf{r},\mathbf{r'})$ is a genuine three-dimensional Coulomb 
potential including the effects of the two semi-infinite metallic electrodes.
The electrostatics of the molecular junction is then governed by this 
potential. It remains to construct and solve the Schr\"odinger equation for the 
molecular wire.

\subsection{Tight-binding model including image charges}

In the following, we will mainly be concerned with conjugated molecules and,
in particular, with their low-energy properties. An appropriate description
can then be given by an effective tight-binding Hamiltonian for the $\pi$
electrons only \cite{salem,baeriswyl}. In addition, since there exist
\textit{ab initio} results for short chains of gold atoms \cite{damle}, it
is also of interest to study systems with electrons in s-orbitals. 

We therefore attach Gaussian type orbitals of the form
\begin{equation}
\label{gaussian}
\phi_n(\mathbf{r})=A_sx^s\exp\big[-\alpha [x^2+y^2+(z-z_n)^2]\big]
\end{equation}
to each atomic site $n$. This allows us to model both s-orbitals with $s=0$ 
and p-orbitals with $s=1$. The center of the orbital, $z_n=d+(n-1)a$, depends 
on the distance $d$ between each electrode and the molecule as well as 
the lattice constant $a$ in the molecule (cf.\ Fig.~\ref{fig1}). The parameter 
$\alpha$ determines the spread of the state and gives an estimate of the 
electronic density. Finally, 
the normalization constants for s- and p-orbitals are given by 
$A_0=(2\alpha/\pi)^{3/4}$ and $A_1=2(2/\pi)^{3/4}\alpha^{5/4}$, respectively. 

The explicit form (\ref{gaussian}) of the orbitals allows us to determine the 
effective parameters entering the tight-binding model. In the following, we adopt the `Zero Differential Overlap' (ZDO) approximation \cite{salem,baeriswyl,fulde}
\begin{equation}
\label{zdo}
\phi_n^{\star}(\mathbf{r})\phi_m(\mathbf{r})=|\phi_n(\mathbf{r})|^2\delta_{n,m}
\end{equation}which remains valid as long as the orbitals $\phi_n$ are strongly localized on the atomic sites $n$. It implies orthogonality between the Gaussian orbitals on different sites and, most importantly, leads to a drastic reduction of non-vanishing matrix elements of the Coulomb operator since only the two center integrals are retained in the final model. In particular, no exchange integrals will appear. These integrals, involving a differential overlap, are usually negligibly small compared to the Coulomb integrals. Moreover, we approximate the
positive cores by point charges localized at the atomic sites $m$ with
coordinates $(0,0,z_m)$. Then, the energy of an electron localized at
site $n$ due to all positive core charges and their images becomes
\begin{equation}
\label{onsite}
\epsilon_n=-U\sum_{m=1}^N \int d^3\mathbf{r} |\phi_n(\mathbf{r})|^2
\varphi(\mathbf{r},z_m\mathbf{\hat{z}})\;.
\end{equation}
Within the ZDO approximation the only finite Coulomb matrix elements are related to the interaction energy between electrons localized at
sites $n$ and $n'$
\begin{equation}
\label{coulombterm}
U_{n,n'}=U\int d^3\mathbf{r} \int d^3\mathbf{r'} |\phi_n(\mathbf{r})|^2
\varphi(\mathbf{r},\mathbf{r'})|\phi_{n'}(\mathbf{r'})|^2\;.
\end{equation}
It is worthwhile to notice that the interaction terms (\ref{onsite}) and 
(\ref{coulombterm}) depend on the position $n$ along the chain due to the 
image charges but also because of the finite size of the molecular wire.

Within the ZDO approximation kinetic energy contributions vanish as a consequence of (\ref{zdo}). To lowest order, the overlap leads to a constant shift of the on-site energy $\epsilon_n$ that may be disregarded and to nearest neighbor hopping. The hopping matrix element $t$ cannot be evaluated directly within the ZDO and, therefore, has to be treated as a parameter of the model \cite{salem,baeriswyl,fulde}. However, it is possible to relax the approximation \cite{fulde} and estimate the hopping matrix elements from (\ref{gaussian}). Doing so, we have found that their dependences on $n$ are not pronounced, 
and, moreover, this kind of more sophisticated treatment would not change qualitatively our final conclusions. 
Therefore, we assume the hopping matrix elements, $t$, to be constant along the chain, and use the ratio $U/t$ as a parameter to examine the importance of electron-electron interaction.

With the parameters just discussed, we obtain a description of the electrons
in terms of an effective tight-binding model which includes long range 
Coulomb interaction \cite{salem,baeriswyl}
\begin{align}
H &= \sum_{n,\sigma}(\epsilon_n+v_n) c^{\dagger}_{n,\sigma} c_{n,\sigma}
+\sum_{n,\sigma}t(c^{\dagger}_{n+1,\sigma} c_{n,\sigma}+\mathrm{h.c.})
\nonumber\\
&\quad+\frac{1}{2}\sum_{n,n',\sigma,\sigma'}U_{n,n'}c^{\dagger}_{n,\sigma} 
c^{\dagger}_{n',\sigma'} c_{n',\sigma'}c_{n,\sigma}\;.
\label{tb}
\end{align}
$c^{\dagger}_{n,\sigma}$ ($c_{n,\sigma}$) are the usual creation (annihilation) operators for an electron with spin $\sigma$ in the local state $\phi_n$. In the first term, we have accounted for an additional shift of the local 
potential due to an external bias. With the chemical potentials $\mu_l$ and 
$\mu_r$ in the left and right electrode, respectively, the shift at site $n$ 
is given by
\begin{equation}
\label{rampe}
v_n=\mu_l+\frac{\mu_r-\mu_l}{L}z_n\;.
\end{equation}

The resulting tight-binding model (\ref{tb}) is mostly defined by the geometry 
of the molecular junction. The only other parameters are the energies $t$ and 
$U$, which determine the strength of the kinetic and Coulomb energies, 
respectively. It is usually believed that conjugated molecules lie in an 
intermediate regime where $U/t=1\dots4$ \cite{salem,baeriswyl}.

\section{Electron density without bias potential}
\label{sec:nobias}

Consider first the situation without bias potential, $\mu_l=\mu_r$ or 
$v_n=0$. We analyse for this case the electron density at equilibrium in the 
absence of electron transfer between the molecule and the leads. In all the calculations presented here, we assume the 
chains to be electrically neutral with on average one electron per site.
Because of the electron spin, this corresponds to the half-filled case. Furthermore,
we restrict ourselves to the ground state of the system. 

\begin{figure}
\includegraphics[width=\linewidth]{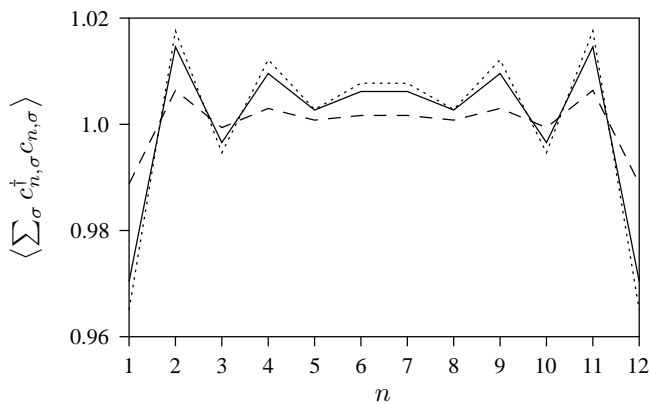}
\caption{Electronic density without bias potential for a molecular wire of 
carbon ($s=1$) with $N=12$ sites. The parameters are $\alpha=4.5/a^2$ and $U=t$. 
The full curve is the exact result for a molecule-electrode distance of $d=2a$. 
The dotted curve is for the same geometry but at Hartree-Fock level. The 
dashed curve is the exact result for the case without metallic electrodes, 
$d\rightarrow \infty$.}
\label{fig2}
\end{figure}

We have done exact diagonalization studies for chains of up to 12 sites. 
Results for the charge density of a chain with 12 sites shown in Fig.~\ref{fig2} are typical for other cases studied. Two main 
features can easily be seen. (i) The electron density is non-uniform: the 
electrons have a tendency to shift towards the middle of the chain. (ii) 
Because of this non-uniformity, substantial Friedel oscillations occur across 
the wire. These features can be explained by invoking electron-hole symmetry as 
it is shown below.

For the case of a half-filled band it has long been known that the electron 
density is uniform, i.e.\ does not depend on the site
index $n$, for models defined on bipartite lattices in such a way that 
electron-hole symmetry is fulfilled. This theorem was first discovered for the 
ground state of the free-electron (H\"uckel) model \cite{coulson} and later 
extended to some interacting systems \cite{maclachlan}. More recently, a
generalization to canonical and grand canonical ensembles was proven for a 
large class of models \cite{lieb}. Mathematically, the theorem applies to 
models invariant with respect to the transformation 
$c^{\dagger}_{n,\sigma}\rightarrow (-1)^n c_{n,\sigma}$.

Applying this transformation, we find that the Hamiltonian of the molecular 
wire (\ref{tb}) is invariant only if the equality
\begin{equation}
\label{uniformity}
\epsilon_n=-\frac{1}{2}\sum_{m=1}^NU_{n,m}\left(1-\frac{\delta_{n,m}}{2}\right) + K
\end{equation} 
is fulfilled where $K$ is a constant. This implies that the 
electron-ion interaction is essentially compensated, up to a constant term, by the repulsive electron-electron interaction. Notice that allowing for hopping matrix elements that depend on the position in the chain would not change this equality. Indeed, a term like $\sum_{n,\sigma} t_{n,n+1}(c^{\dagger}_{n+1,\sigma}c_{n,\sigma}+c^{\dagger}_{n,\sigma}c_{n+1,\sigma})$ remains unchanged when the electron-hole transformation is applied. From this point of view, it is not necessary to go beyond the ZDO approximation.

The equality (\ref{uniformity}) can be satisfied only in very particular cases
where at least one of the following conditions is satisfied:
\begin{itemize}
\item the chain is infinite;
\item there is no interaction, $U=0$;
\item $\alpha \rightarrow \infty$, corresponding to interactions between point 
charges.
\end{itemize}

None of these criteria are fulfilled in realistic cases of interest. With the exception of carbon nanotubes, experimental molecular junctions
involve relatively short ($<10\,\mathrm{nm}$) molecular chains \cite{joachim}. 
The Coulomb interaction is of the same order of magnitude as the kinetic terms 
\cite{salem,baeriswyl} and certainly not negligible. Even then, the 
electron density would still be uniform if the electron-orbitals are reduced to 
points ($\alpha \rightarrow \infty$) irrespective of the presence of image 
charges. Therefore, one may say that the non-uniformity of the density comes 
from the lateral extension of the electronic clouds, the p- or s-states. This 
shows, once again, the need of a three-dimensional treatment of the electronic 
structure.

The exact shape of the electron density is determined by (i) the strength of 
the Coulomb interaction, $U/t$, (ii) the spread of the electronic orbitals, 
$\alpha$, and (iii) the geometry of the system, expressed in our model by
the length $(n-1)a$ of the molecule and by the distance $d$ between the 
electrodes and the molecule.

All these factors contribute to yield, instead of the uniform density condition 
(\ref{uniformity}), the relation
\begin{equation}
\epsilon_n=-\frac{1}{2}\sum_{m=1}^NU_{n,m}\left(1-\frac{\delta_{n,m}}{2}\right) + f_L(n)
\end{equation}
where $f_L(n)$ is a function of the position in the chain; its dependence on 
$n$ is responsible for the non-homogeneity of the electronic density. It is of 
interest to understand the effects of each of these parameters
separately.

{\it Coulomb interaction}. Starting from the non-interacting case where the 
electronic density is uniform, increasing $U$ results in an increase of the 
Friedel oscillations until the electrons start to localize. In the limit
$U\to\infty$, half filling leads to a Mott insulator and one recovers a
constant electron density.

{\it Spread of the electronic orbitals}. When the orbitals are reduced to 
points (point charge limit), $\alpha \rightarrow \infty$, the electronic 
density is uniform. Indeed, in this particular limit the electron-ion and electron-electron interactions have the same form restoring the electron-hole symmetry of the molecular Hamiltonian. Accounting for a spread of the orbitals increases slowly 
the amplitude of the Friedel oscillations.

{\it Geometry of the molecular junction}. Two effects need to be distinguished: 
(i) the finite size effects and (ii) the image-charge effects. 
In Fig.~\ref{fig2}, 
the dashed curve shows the density for the very same system used to obtain the 
other two curves except that now no metallic electrodes are present. Therefore, 
the dashed curve contains only the finite size effects. From this 
particular example, one sees that both effects contribute to the Friedel 
oscillations and that, in order to get the correct density, image interactions should be included unless the electrode-molecule distances is large. In fact, in the example of Fig.~\ref{fig2}, the contribution to 
the oscillations due to the image charges is the larger one. The Coulomb 
interaction with the electrodes described by the image charges is therefore 
crucial to estimate transport properties of molecular wires or metallic 
constrictions. We have observed, as expected, that finite size effects alone 
tend to disappear when the chain size is increased. In contrast, the effect
of image charges tends to become more important: the amplitude of the Friedel oscillations 
increases with the system size due to the presence of the two metallic 
electrodes. This tendency should continue until for long wires, that we do not 
consider here, the oscillations occur predominantly near the edges. 

Finally, the dotted curve of Fig.~\ref{fig2} shows the electron density with 
metallic electrodes calculated within the Hartree-Fock approximation; it is in 
very good agreement with the exact result. We have performed calculations for 
different values of $U$ up to $U=4t$ and observed that, as far as the electron 
density is concerned, the Hartree-Fock approximation gives reasonable results. 
This allow us to use this mean-field approach for the study of the 
electrostatic profile in biased molecular wires (section~\ref{sec:bias}) and 
their transport properties (section~\ref{sec:iv}).

\section{Electrostatic potential in biased molecular wires}
\label{sec:bias}

The application of the screened on-site electrostatic potential to
calculations of conductance properties is useful only within a single electron theory, e.g. in
the mean-field description of the process. This is how this concept was
applied in the calculation of Mujica \textit{et al.}\ \cite{mujica1} and in
the density functional theory \cite{lang,damle}. Here, our calculations are 
done using the Hartree-Fock approximation applied to the molecular 
Hamiltonian (\ref{tb}). In the previous section, by comparing the mean-field electron density with the exact one, we have already shown, that the approximate result is reasonable in the range of parameters proper to conjugated 
molecules. More precisely, the Hartree-Fock density is in very good agreement 
with the exact results for values of $U$ up to, approximately, $2t$ 
(cf.\ Fig.~\ref{fig2}) and remains of reasonable accuracy up to values of about $4t$. In the following we present 
only results for $U=t$, but the same qualitative picture arises also for larger 
values of $U$.  

Starting from the initial tight-binding model (\ref{tb}), we build an effective 
Hamiltonian by solving self-consistently the usual Hartree-Fock equations 
\cite{salem} leading to
\begin{equation}
\begin{aligned}
H_{\rm eff} &=\sum_{n,\sigma}\epsilon_n(V)c^{\dagger}_{n,\sigma}c_{n,\sigma}\\
&\quad+\sum_{n,n',\sigma}t_{n,n'}(V)(c^{\dagger}_{n',\sigma}
c_{n,\sigma}+\mathrm{h.c.})
\end{aligned}
\end{equation}
where $V=\mu_l-\mu_r$. The chemical potentials will always be chosen such
that $\mu_l=-\mu_r$. $\epsilon_n(V)$ is the effective on-site potential which 
includes the ionic attraction and the electron-electron repulsion incorporated 
within a mean field picture as well as the on-site potential (\ref{rampe}). 
$t_{n,n'}(V)$ is the effective hopping matrix element which includes the 
exchange terms. Note that a vanishing of exchange integrals in the ZDO 
approximation is not equivalent to an Hartree approximation. Indeed, a mean 
field approximation of (\ref{tb}) still contains exchange terms of the form 
$-\frac{1}{2}\sum_{n \ne n'}U_{n,n'}\langle c^{\dagger}_{n,\sigma}c_{n',\sigma}
\rangle c^{\dagger}_{n',\sigma}c_{n,\sigma}$ which, due to the long range part 
of the Coulomb potential (\ref{coulomb}), include long range hopping.

Without Coulomb interactions, the electrostatic potential is given by the ramp 
defined in (\ref{rampe}), i.e.\ by the potential in the absence of a molecule. 
This linear profile has been used sometimes in the literature to study 
non-linear current-voltage characteristics \cite{mujica2,mazumdar}.

In the presence of Coulomb interaction, the linear profile is changed and the 
screened electrostatic potential, $E_n(V)$, is given by the difference 
between the on-site term with and without bias voltage. 
\begin{equation}
E_n(V)=\epsilon_n(V)-\epsilon_n(0)
\end{equation}

\begin{figure}
\includegraphics[width=\linewidth]{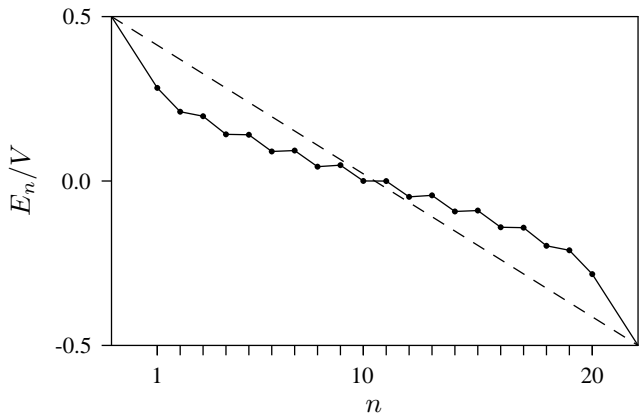}
\caption{Electrostatic potential profile for a carbon wire ($s=1$ in Eq.\ 
\ref{gaussian}) with $N=20$ sites. The parameters are $\alpha=4.5/a^2$, $U=t$, 
$V=\mu_l-\mu_r=t$ and $d=2a$. The dashed curve is the unscreened potential 
without molecule (ramp potential). The full curve is the screened potential in 
the presence of the molecular wire. It shows a small decrease in slope along 
the entire molecule with substantial Friedel oscillations.}
\label{fig3}
\end{figure}

A typical example is shown in Fig.~\ref{fig3} for a chain with 20 sites. We see two main features: (i) screening is not very efficient and, consequently, 
there is only a small potential drop at the interfaces but a finite slope of the potential along the 
entire molecule; (ii) there are substantial Friedel oscillations along the 
profile.

\begin{figure}
\includegraphics[width=\linewidth]{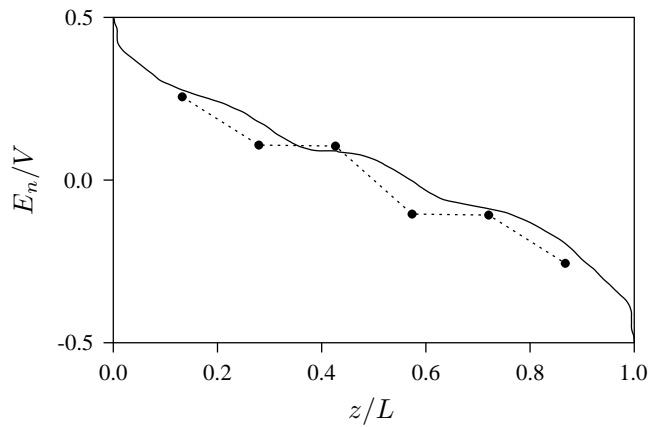}
\caption{The electrostatic potential profile for a gold wire ($s=0$ in 
Eq.~\ref{gaussian}) with $N=6$ sites obtained from Hartree-Fock calculations 
is shown by the filled circles. The parameters used are $d=0.9a$, 
$\alpha=4.5/a^2$, and $U=t$. For comparison the full curve shows the
\textit{ab initio} results taken from Ref.~~\onlinecite{damle}.}
\label{fig4}
\end{figure}

Our results differ from those of Mujica \textit{et al.}\ \cite{mujica1} 
despite the fact that the two models are closely related. These differences 
mostly come from the fact that we solve the Poisson equation without 
resorting to a one-dimensional approximation. As recently shown by us
within a classical model \cite{nitzan2002}, a three-dimensional 
treatment of the electrostatics is necessary, and in fact leads to the
identification of the lateral thickness of the molecule as a new generic
attribute that determines the potential profile. In Fig.~\ref{fig4}, we show a comparison between 
our results and the calculations of Damle \textit{et al.}\ \cite{damle} for a 
chain of six gold atoms. It is important to stress that we did not try to fit 
the \textit{ab initio} curve but, instead, we simply chose a reasonable set of 
parameters. Note, however, that we include the response of infinite leads while the \textit{ab initio} calculations take only small metal clusters into 
account. Furthermore, the asymmetry in the latter case indicates that the
\textit{ab initio} calculations result in a charged molecule while in our case 
the molecular wire always remains neutral. Other \textit{ab initio} 
calculations \cite{lang,weber,damle2002} on similar models are also in 
agreement with our observations.  

To conclude this section, we summarize our main results: in the relevant range 
of parameters, screening is not very efficient in the wire and the drop of the 
potential occurs along the entire molecule. Additionally, substantial Friedel 
oscillations are present in the electrostatic profile. Our results are in good 
agreement with recent \textit{ab initio} results \cite{lang,damle}. When 
compared with the results of Ref.~~\onlinecite{mujica1}, our treatment stresses 
the need to study the full three-dimensional problem, as far as the 
electrostatics of the molecular junction is concerned. 

In the next section, we discuss consequences of the screening effects for the current-voltage characteristics of molecular wires studied within the Landauer formalism.

\section{Current-voltage characteristics of a molecular wire in the weak 
tunnelling contact limit}
\label{sec:iv}

In a model where only the first and the last atom of a molecular chain
couple to the corresponding metal leads and at $T=0$, the Landauer
conduction formula yields \cite{bardeen,nitzan}
\begin{equation}
\label{current}
I=\frac{2e}{\pi\hbar}\int^{\mu_l}_{\mu_r}dE|G_{1N}(E,V)|^2\Delta_l(E,V)
\Delta_r(E,V)
\end{equation}
where $V=\mu_l-\mu_r$ and $\Delta_{l/r}$ are the spectral functions for the left 
and right reservoirs. The molecule-lead coupling only occurs at sites $1$ and $N$ and $G_{1N}$ is the matrix element of the exact Green 
function of the molecular junction between these sites (cf.\ Fig.~\ref{fig1}).

This equation can be understood as a special case of the Landauer formula 
\cite{landauer} adapted to the case of bad contacts, where it is possible to 
use second order perturbation theory in the tunnelling matrix element 
\cite{bardeen}. In the limit of `extremely' bad contacts of interest here, the Green function of the system in formula (\ref{current}) may be 
replaced by the Green function of the isolated molecular wire, $G^0_{1N}$ 
\cite{onipko}. Moreover, assuming that the product of spectral densities does 
not significantly depend on energy in the range between $\mu_l$ and $\mu_r$, one gets
\begin{equation}
\label{current0}
I=\frac{2e}{\pi\hbar}\Delta_0^2\int^{\mu_l}_{\mu_r}dE|G_{1N}^0(E,V)|^2
\end{equation}
where $\Delta_0$ is the spectral density at zero bias.

\begin{figure}
\includegraphics[width=\linewidth]{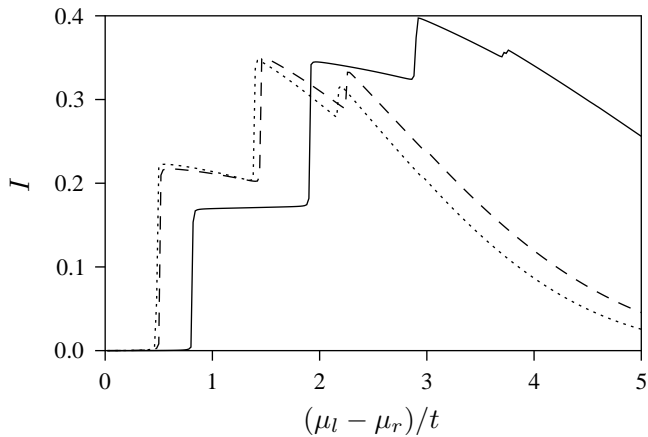}
\caption{Current-voltage characteristics for a carbon molecule ($s=1$) with 
$N=12$, $d=2a$, $\alpha=4.5/a^2$, and for different Coulomb interaction 
strengths $U/t=0, 0.1, 1$ are depicted by the dotted, dashed, and full line, 
respectively. The current is given in units of $(2e/\pi\hbar)\Delta_0^2/t$.} 
\label{fig5}
\end{figure}

From this simplified equation, we can calculate the current-voltage 
characteristics of the molecular junction, using only the 
Green function of the isolated molecular wire, evaluated in the presence of the electrostatic 
potential created by the metallic electrodes. Examples are given in
Fig.~\ref{fig5} for increasing Coulomb strength $U$ starting from the 
non-interacting case where the on-site potential is given by the ramp (\ref{rampe}).

In all cases, the $I$-$V$ curves have a staircase structure which is a common 
feature in the weak tunnelling limit \cite{damle,mujica2,foot} and simply reflects 
the discreteness of the molecular electronic spectrum. Indeed, an increase of 
the bias potential corresponds to an increase of the window of integration in 
formula (\ref{current0}). Therefore, a jump in the $I$-$V$ curves means that 
one more discrete molecular level enters this window of integration.

It is interesting to note that Fig.~\ref{fig5} shows also wide regions of 
negative differential resistance, in particular in the non-interacting case. 
Within our simple formulation, they can be explained by the localization of 
charges induced by a strong enough electric field: at sufficiently high bias 
voltage, charge carriers are localized at one end of the chain resulting in a 
decrease of the current. This could be an artifact of our weak molecule-electrode coupling model, however negative differential resistance has been found in 
recent experiments \cite{chen}, and electric field induced localization could 
be a way to understand these experimental findings.

The Coulomb interaction has two main effects on the $I$-$V$ characteristics. 
First, the positions of the current-voltage steps are shifted to higher 
voltages reflecting the displacement of the molecular levels to higher energy 
with increasing Coulomb interaction. Second, the localization of the charges 
due to a strong electric field is partially compensated by the 
electron-electron repulsion. These screening effects attenuate the negative 
differential resistance effects (cf. Fig.~\ref{fig5}). These observations are similar to those made by Mujica et al.\ \cite{mujica2}, 
where a Hubbard model was studied at the Hartree level.

\section{Conclusion}

We have addressed the problem of calculating transport properties of a 
molecular wire bridging two semi-infinite metallic electrodes. A first 
important part of this task is to determine the electrostatic potential profile 
through the biased wire \cite{datta,tian,mujica1,lang,damle}. Indeed, it is of 
importance to know how screening effects modify the ramp potential 
(Eq.~\ref{rampe}) existing in the absence of the organic molecule.

This work and our earlier paper \cite{nitzan2002} resolve discrepancies between 
answers available in the literature. On the one hand, a tight-binding model 
combined with a one-dimensional Poisson equation gives a strong screening 
version of the problem: the drop of the potential occurs at the interfaces and 
the potential is therefore almost flat within the molecule \cite{mujica1}. On 
the other hand, recent \textit{ab initio} results give a weak screening 
version: no substantial drop at the interfaces but rather a decrease along the 
entire molecule together with substantial Friedel oscillations 
\cite{lang,damle}.

In this paper, we have proposed a modified tight-binding model to address this 
question. It is based on three main ingredients. (i) We introduce a 
three-dimensional Coulomb potential (Eq.~\ref{coulomb}) which includes the 
image interaction with the two metallic electrodes assumed to have planar surfaces. 
(ii) The electrons localized at atomic sites are modelled by Gaussian type 
orbitals of finite lateral extent (Eq.~\ref{gaussian}). (iii) The positive 
background is assumed to be a set of point charges localized on the atomic 
sites. With these three ingredients, it is possible to evaluate the 
various terms of our model Hamiltonian: the on-site energy (Eq.~\ref{onsite}) 
and the electron-electron interaction (Eq.~\ref{coulombterm}). They are all 
functions of the position on the chain due to the finite size of the system and 
the image charges induced by the electrodes. All our calculations are done in 
the weak tunnelling limit, assuming bad contacts.

This model yields an electronic density that is non-uniform already in the absence of a bias 
potential --- the electrons prefer to be in the center of the wire --- and 
displays pronounced Friedel oscillations (Fig.~\ref{fig2}). These 
characteristics are due to the fact that our model does not fulfill, in 
general, the electron-hole symmetry \cite{coulson,maclachlan,lieb}. In the 
presence of an applied voltage the electrostatic potential profile does not
drop appreciably  at the interfaces but rather, in accordance with the
\textit{ab initio} results of Refs.~~\onlinecite{lang} and\ \ \onlinecite{damle}, 
it decreases 
along the molecule with substantial Friedel oscillations appearing along the 
entire profile (Figs.~\ref{fig3} and \ref{fig4}). Our results are different 
from the ones of Ref.~~\onlinecite{mujica1} where a one-dimensional Poisson 
equation was used. This disagreement stresses the need to perform a 
three-dimensional calculation, as done here, to properly describe the 
electrostatic properties of the molecular junction. 
Finally, the current-voltage characteristics shows a staircase 
structure (Fig.~\ref{fig5}) as is common in the weak tunnelling limit 
\cite{damle,mujica2}. Zones of negative differential resistance are found due 
to charge localization induced by the electric field. The main effects of the 
Coulomb interaction, within our approximations, are, on the one hand, a shift 
to higher energies of the position of the current-voltage steps and, on the 
other hand, a partial compensation of the localization of the electrons 
diminishing the negative differential resistance effects in agreement with a 
previous study \cite{mujica2}.

In closing, it is important to stress some limitations of our model. On the one hand, we consider coherent transport, assuming that the electrons are transferred from one lead to the other in a single quantum mechanical process. This is a good approximation if the tunnelling time is much less than the inelastic scattering time. For organic molecules, this transit time could be of the same order of magnitude as the intramolecular vibronic relaxation time, especially in the weak contact limit employed here \cite{nitzan}. In this case, part of the current could be due to sequential tunnelling, where the molecular wire would be successively charged and discharged. This important issue remains to be studied further. We have neglected charging effects assuming the molecule to remain neutral. At high voltage, this approximation could fail \cite{mujica2}. The average charge number of the molecule could increase in analogy with Coulomb blockade phenomena observed in mesoscopic metallic double tunnel junctions and quantum dots \cite{grabert}, and, more recently, in multiwall carbon nanotubes \cite{buitelaar}.

A proper handling of the full problem requires to treat the wire as an open system dynamically coupled to the electrodes and to the vibronic degrees of freedom. This program is far beyond the scope of the present work.

\begin{acknowledgments}
The authors have benefitted from useful discussions with P.~H{\"a}nggi,
S.~Kohler, J.~Lehmann, and S.~Yaliraki. Three of us (HG, GLI and AN) would like
to thank the Institute for Theoretical Physics at UCSB for hospitality
during the workshop on ``Nanoscience'' where this work was started. This
research was supported in part by the National Science Foundation under
Grant No.\ PHY99-07949, by the Volkswagen-Stiftung under grant No.\
I/77\,217, by the Israel Science Foundation and by the Israel Ministry of
Science.
\end{acknowledgments}


\begin{thebibliography}{}
\bibitem{foot}
A finite molecule-lead coupling matrix element would smoothen the sharp steps in Fig. \ref{fig5}.
\end{thebibliography}
\end{document}